\documentclass[newstyle,twocolumn,journal]{rmaa}
\usepackage{paralist}

\usepackage{psfrag,color}

\usepackage[latin1]{inputenc}

\title{The inclination of the dwarf irregular galaxy 
Holmberg II}
\author{F.~J.~S\'anchez-Salcedo\altaffilmark{1},
          A.~M.~Hidalgo-G\'amez\altaffilmark{2} and 
          Eric E.~Mart\'{\i}nez-Garc\'{\i}a\altaffilmark{3}}

\altaffiltext{1}{Instituto de Astronom\'\i a, Universidad Nacional Aut\'onoma
de M\'exico, Mexico City, Mexico }
\altaffiltext{2}{Departamento de F\'{\i}sica, Escuela Superior de F\'{\i}sica
y Matem\'aticas, IPN, Mexico City, Mexico}
\altaffiltext{3}{Instituto Nacional de Astrof\'{\i}sica, \'Optica y Electr\'onica
(INAOE), Puebla, Mexico}

\shortauthor{S\'ANCHEZ-SALCEDO ET AL.}

\shorttitle{The inclination of Ho II}

\fulladdresses{
\item F. J. S\'anchez-Salcedo: Instituto de Astronom\'{\i}a, Universidad Nacional Aut\'onoma
de M\'exico, P.O. Box 70-264, Ciudad Universitaria, C.P. 04510,
Mexico City, Mexico (jsanchez@astro.unamm.mx).
\item A. M. Hidalgo-G\'amez: Departamento de F\'{\i}sica, Escuela Superior de F\'{\i}sica
y Matem\'aticas, Instituto Polit\'ecnico Nacional, U.P. Adolfo L\'opez Mateos,
C.P. 07738, Mexico City, Mexico.
\item E. E. Mart\'{\i}nez-Garc\'{\i}a: Instituto Nacional de Astrof\'{\i}sica, \'Optica y Electr\'onica
(INAOE), Apartado Postal 51 y 216, C.P. 72000, Puebla, Mexico.
}

\listofauthors{F.~J.~S\'anchez-Salcedo, A.~M.~Hidalgo-G\'amez, \&
E. E. Mart\'{\i}nez-Garc\'{\i}a}
\indexauthor{S\'anchez-Salcedo, F.~J.}
\indexauthor{Hidalgo-G\'amez, A.~M.}
\indexauthor{Mart\'{\i}nez-Garc\'{\i}a, E. E.}
\resumen{
Damos restricciones
al \'angulo de inclinaci\'on del disco de H\,{\sc i} de la galaxia irregular
enana Holmberg II (Ho II) a partir de un an\'alisis de la estabilidad gravitacional del
disco de gas en las partes externas.  Encontramos que se requiere un \'angulo de inclinaci\'on medio de
$27^{\circ}$ y, por lo tanto, una velocidad circular en su parte plana de $\sim 60$ km s$^{-1}$, para
tener un grado de estabilidad similar al que tienen otras galaxias. Para esa inclinaci\'on,
Ho II cae en la posici\'on correcta en la relaci\'on de Tully-Fisher bari\'onica
y, adem\'as, su curva de rotaci\'on es consistente con MOND. 
Sin embargo, el an\'alisis de estabilidad
correspondiente indica 	que esta galaxia podr\'{\i}a ser problem\'atica para MOND
porque las partes externas de
esta galaxia ser\'{\i}an marginalmente inestables bajo esa teor\'{\i}a de gravedad.
Se requieren simulaciones num\'ericas de las galaxias enanas
ricas en gas para ver la factibilidad de MOND.
}

\abstract{ 
We provide constraints on the inclination angle of the H\,{\sc i} disk of the
dwarf irregular galaxy Holmberg II (Ho II) from
stability analysis of the outer gaseous disk. 
We point out that a mean inclination angle of $27^{\circ}$ and thus
a flat circular velocity of $\sim 60$ km s$^{-1}$, is required to have
a level of gravitational stability similar to that found in other galaxies. 
Adopting this inclination angle, we find that Ho II lies on the right
location in the baryonic Tully-Fisher relation. Moreover,
for this inclination, its rotation curve is consistent with MOND.
However, the corresponding analysis of the stability under MOND indicates that 
this galaxy could be problematic for MOND because its outer parts are
marginally unstable in this gravity theory. We urge MOND simulators to 
study numerically the non-linear stability of gas-rich dwarf galaxies since
it may provide a new key test for MOND.}

\addkeyword{dark matter}
\addkeyword{galaxies: individual (Holmberg II)}
\addkeyword{galaxies: kinematics and dynamics}
\addkeyword{gravitation}

\begin{document}
\maketitle

\section{Introduction}
The gas rotation curves of spiral galaxies are useful to derive the
radial distribution of their dynamical mass and to test modified
gravity theories as alternatives to dark matter. Accurate determinations 
of the rotation curves of low-surface brightness galaxies and dwarf 
irregular galaxies are very useful to test the predictions of the numerical 
simulations of structure formation (see de Blok 2010 for a review). In these galaxies, it is
possible to derive the dark matter profiles also in their central parts 
because the contribution of the dark matter to the
potential is dominant at any galactocentric radius.

Empirical correlations between the value of the circular velocity in the flat 
part ($v_{\rm flat}$) and other
global properties of the galaxy (e.g., the baryonic mass)
provide strong constraints to the theories of formation and
evolution of galaxies (e.g., Silk \& Mamon 2012). 
In order to derive the amplitude of the H\,{\sc i} rotation curve of a certain
galaxy, a good estimate of its inclination angle in the sky, $i$, is needed 
to deproject the observed velocity field. The inclination can be
determined from the morphology or by modeling the entire
two-dimensional radial velocity field
using the tilted-ring analysis (e.g. Begeman et al.~1991).
In irregular galaxies with large scale asymmetries (e.g., stellar bars or 
giant holes in the H\,{\sc i} surface density distribution) and/or
rising rotation curves without a well-defined flat part,
inclination is an uncertainty. 
As an example, consider the
dwarf irregular galaxy IC 2574. Inspection of the optical image
suggests that this galaxy is highly inclined.  Martimbeau et al.~(1994)
derived an inclination angle of $77^{\circ}\pm 3^{\circ}$ from the central
isophotes in the R-band and a mean inclination of $75^{\circ}\pm 7^{\circ}$ from 
the H\,{\sc i} kinematical analysis. Following a similar procedure, Walter \& Brinks (1999) 
confirmed the value of the inclination by Martimbeau et al.~(1994).
More recently,
Oh et al.~(2011), using a more sophisticated analysis to minimize
the effect of non-circular motions, obtained an average value of the
inclination angle of the H\,{\sc i} disk of only $53^{\circ}$. 

Since the rotation velocity depends on the adopted inclination as $1/\sin i$, 
uncertainties in the inclination are more important in galaxies with low
inclinations (i.e.~more face-on galaxies).
In this paper, we suggest an indirect method to give upper limits on $i$ in
galaxies with low inclinations, where traditional methods are less precise.
We will focus on the dwarf irregular galaxy Holmberg II (Ho II) or DDO 50.
From tilted-ring models, Oh et al.~(2011) derived a mean inclination angle
of $49^{\circ}$. For this inclination, the asymptotic (flat) circular velocity 
Ho II
is $36$ km s$^{-1}$, which is inconsistent with MOND (S\'anchez-Salcedo
et al. 2013). Motivated by this result, Gentile et al.~(2012) re-analyzed the data cubes
at the outer parts in Ho II ($R>4$ kpc) and concluded that an inclination angle
between $20^{\circ}$ and $35^{\circ}$ is more consistent with the observed
total H\,{\sc i} map (see also McGaugh 2011).
Here, we invoke stability arguments to provide constraints on the inclination angle
of Ho II and, thus, on the amplitude of its rotation curve.

The paper is organized as follows. Section \ref{sec:properties} reviews some
structural and kinematical parameters of Ho II. In \S \ref{sec:newcase}, we compute
the Toomre parameter of the gaseous disk and compare it with the values
obtained in other galaxies.  Section \ref{sec:normality} shows how the Toomre parameter
changes if a different inclination angle is adopted.  We suggest that a mean inclination
angle of the H\,{\sc i} disk of $i\sim 27^{\circ}$ leads Ho II to a  ``normal'' level of stability, 
and it is consistent with the baryonic Tully-Fisher relation (e.g, McGaugh 2012), with 
the Zasov-Smirnova relation (Zasov \& Smirnova 2005) and with other empirical trends.
The implications of adopting this lower inclination for MOdified Newtonian 
Dynamics (MOND) are discussed in \S \ref{sec:MONDcase}.
Conclusions are given in \S \ref{sec:conclusions}.

\section{Global properties of Ho II}
\label{sec:properties}
Ho II (DDO 50) is a relatively nearby gas-rich dwarf irregular galaxy in the M81 group.  
Based on the tip of red giant branch method,  Karachentsev et al.~(2002) derived a distance
of $3.4$ Mpc, whereas Hoessel et al.~(1998) inferred a distance of $3.05$ Mpc using
Cepheids.
To facilitate comparison with the works of Leroy et al.~(2008) and Oh et al.~(2011), we will
adopt a distance of $3.4$ Mpc but the implications of using a shorter distance are also 
discussed in Section \ref{sec:normality}.
 At a distance of $3.4$ Mpc, it has a total absolute B magnitude of $-16.9$ mag and
a radius  measured to the $25$ B mag arcsec$^{-2}$
surface brightness level, denoted by $R_{25}$, of $3.3$ kpc (Leroy et al. 2008).
The optical inclination, extracted from HYPERLEDA,
is $45^{\circ}$ (Paturel et al.~2003). 

The H\,{\sc i} distribution and its kinematics have been studied by 
Puche et al.~(1992), Bureau \& Carignan (2002), and by Oh et al.~(2011).
There is no consensus on the H\,{\sc i} inclination angle; ellipse fits to the outer
H\,{\sc i} disk (at radius $\sim 7$ kpc) suggests an inclination of $\sim 30^{\circ}$ 
(de Blok et al.~2008;
McGaugh 2011; Gentile et al.~2012), whereas tilted-ring analysis indicates 
an inclination (at $\sim 7$ kpc), roughly between $40^{\circ}$ and $50^{\circ}$
(Bureau \& Carignan 2002; Oh et al.~2011). For $\left<i\right>=49^{\circ}$, the asymptotic
velocity of Ho II is $36$ km s$^{-1}$ (Oh et al.~2011).

\section{The Toomre stability parameter of the gas disk in Ho II}
\label{sec:newcase}
There exist several studies aimed to test the importance of the gravitational
instability in determining both the sites of massive star formation
and the star formation rate in galaxies (Kennicutt 1989; van Zee et al.~1997;
Hunter et al.~1998;
Martin \& Kennicutt 2001; Kim \& Ostriker 2001, 2007; de Blok \& Walter 2006; Yang et al.~2007; 
Leroy et al.~2008; Yim et al.~2011; Elmegreen 2011).
From observations of nearby Sc galaxies, Kennicutt (1989) and 
Martin \& Kennicutt (2001) found that there exists a gas surface density
threshold for star formation. These authors suggested that the threshold
depends on the Toomre parameter of the
gaseous disk, defined as 
\begin{equation}
Q_{g}=\frac{\Sigma_{c}}{\Sigma_{g}},
\label{eq:Qnewton}
\end{equation}
where $\Sigma_{g}$ is the gas surface density (that is, H\,{\sc i}$+$H$_{2}$ gas corrected to include
helium and metals), and
the critical density for instability is
\begin{equation}
\Sigma_{c}=\frac{\kappa c_{s}}{\pi G },
\end{equation}
with $\kappa$ the epicyclic frequency and $c_{s}$ the velocity dispersion of the gas. 
We must emphasize here that Kennicutt (1989) and Martin \& Kennicutt (2001) used the
velocity dispersion of the gas and not the effective sound speed (see Schaye 2004 for
a discussion). Turbulence tends to stabilize the gaseous disk but the magnitude of this
effect is small in disks where the H\,{\sc i} is the main component (Romeo et al.~2010).
In some galaxies, 
the contribution of the stellar disk to gravitational
instability cannot be ignored (Yang et al.~2007). Nevertheless,
in the case of gas-rich dwarf galaxies, as it is the case for Ho II,
or in the outer parts of spiral galaxies, 
the inclusion of the stellar contribution is not so relevant
(see fig.~5 in Hunter et al.~1998; Meurer et al.~2013).

Recent studies indicate that galactic disks stabilize to a constant
stability parameter across the optical galaxy (Meurer et al.~2013; Zheng et al.~2013),
lending support to the phenomenon of self-regulation by
star formation, as suggested by Quirk (1972). However, although the
value of the $Q_{g}$-parameter is approximately constant 
across a galaxy, it varies from galaxy to galaxy, indicating
that the `thermostat'  process (i.e.~the stellar feedback)
is not solely determined by gravitational instabilities
(e.g.~Zheng et al.~2013). In fact, for a sample of spiral galaxies and dwarf
galaxies, Leroy et al.~(2008) found that $Q_{g}$ has a so large scatter
from galaxy to galaxy,
that there is no a clear evidence of a universal $Q_{g}$ threshold marking the
transition from high star formation efficiency to low star formation efficiency
(see also Hunter et al.~1998 and C\^ot\'e et al.~2000).

Hunter et al.~(1998)
first noticed that the dwarf irregular galaxy Ho II presents some peculiarities regarding the values
of the Toomre parameter when compared to other galaxies. Using an inclination angle
of $40^{\circ}$,
they found that beyond a galactocentric radius of $2$ kpc, $\Sigma_{g}/\Sigma_{c}$ in Ho II
is $\sim 3$ times higher than the average peaks in the other
irregular galaxies. Only out at $\sim 7$ kpc, the ratio between
the gas column density and the critical density declines to a value
comparable to the peak values in the other irregular galaxies.

Using the updated high-resolution data for the
rotation curve and the azimuthally averaged H\,{\sc i}
surface density (corrected by a factor $1.4$ to include helium and metals)
for Ho II, as derived in Oh et al.~(2011), 
we have calculated the Toomre parameter $Q_{g}$ and the critical density $\Sigma_{c}$,
for $D=3.4$ Mpc and $c_{s}=6$ km s$^{-1}$. A value for $c_{s}$
of $6$ km s$^{-1}$ corresponds to the value used by Kennicutt (1989). However,
the reason to adopt this value for $c_{s}$  is more profound.
There is growing evidence that suggests that a 
sound speed of $4-6$ km s$^{-1}$ (instead of the observed
velocity dispersion), when used 
to derive the Toomre critical density, gives an optimal description
of ongoing star formation regions (de Blok \& Walter 2006; Yang et al.~2007).
These authors argue that in a multiphase medium,
the measured velocity dispersion is the sum of contributions of
the dispersions of the cool and warm phases with additional
input from star formation and turbulence. As the relevant dispersion
value to use in a star formation threshold analysis is that
of the cool phase of the ISM, using the second-moment values
will likely result in significant overestimates.

\begin{figure}
\includegraphics[width=85mm, height=110mm]{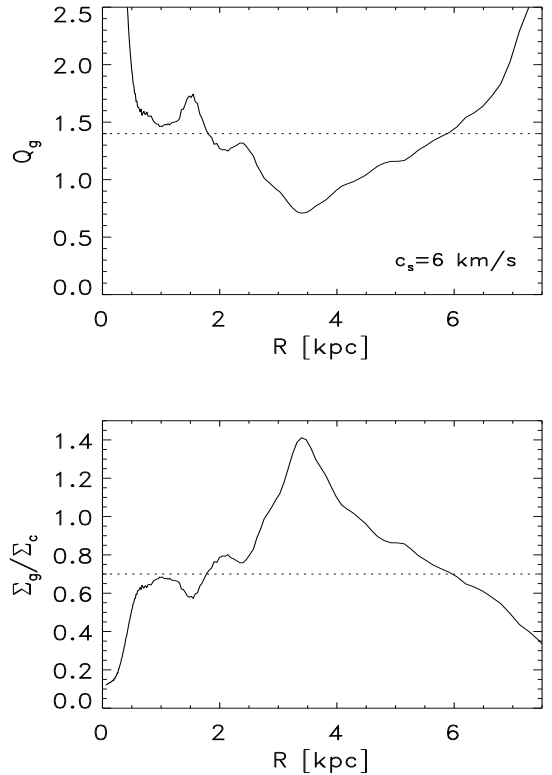}
 \caption{Radial profile of the gas Toomre parameter (top panel)
and $\Sigma_{g}/\Sigma_{c}$ (bottom panel) in Ho II for $c_{s}=6$ km s$^{-1}$,
$D=3.4$ Mpc and
$\left<i\right>=49^{\circ}$. For reference, the horizontal dotted curve corresponds to
$Q_{g}=1.4$ or, equivalently, $\Sigma_{g}/\Sigma_{c}=0.7$.

}
 \label{fig:Qinclination49}
 \end{figure}

\begin{figure}
\includegraphics[width=80mm, height=105mm]{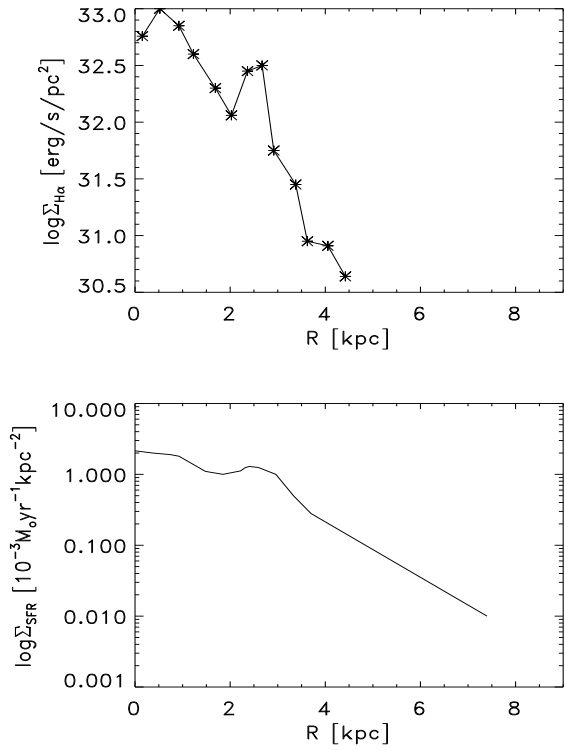}
 \caption{Top panel: Azimuthally averaged H$\alpha$ surface brightness as a function of radius
in Ho II from Hunter et al.~(1998). Bottom panel: star formation rate vs. radius, as derived by
Leroy et al. (2008) inside the optical radius $R_{25}$ (i.e.  where the B-band magnitude
drops below $25$ mag arcsec$^{-2}$) and by Bigiel et al.~(2010) between
$R_{25}$ and $2R_{25}$.

}
 \label{fig:SF}
 \end{figure}

Figure \ref{fig:Qinclination49} shows the radial profile of $Q_{g}$ and
$\Sigma_{g}/\Sigma_{c}$ in Ho II. We see that $Q_{g}<1.5$ in a large
part of the galaxy, from $2$ to $6$ kpc, being minimum at
$3.5$ kpc.  Following Meurer et al.~(2013), it is useful to 
define $R_{1}$ and $R_{2}$ as the radii enclosing $25\%$ and $75\%$ of 
the total H\,{\sc i} gas, respectively. According to Meurer et al.~(2013),
ISM disks do maintain a constant $Q_{g}$ between $R_{1}$ and
$R_{2}$. In the case of Ho II, we find $R_{1}=2.9$ kpc,
$R_{2}=5.5$ kpc, and $\left<Q_{g}\right>=1.0$, where 
$\left<Q_{g}\right>$ is the mean value of $Q_{g}$
between $R_{1}$ and $R_{2}$. 

From theoretical grounds, one would expect a burst of
star formation between $R=2$ and $R=6$ kpc because, as Elmegreen
(2011) showed, dissipative gaseous disks with $Q_{g}<2$--$3$ are
strongly unstable.  Moreover, if Kennicutt's result (1989) is applied, that is,
if the H$\alpha$ emission
extends up to the radius where $Q_{g}\sim 1.4$, then we would predict 
that the H$\alpha$ emission should be detected until $R=6$ kpc.

\begin{figure}
\includegraphics[width=75mm, height=60mm]{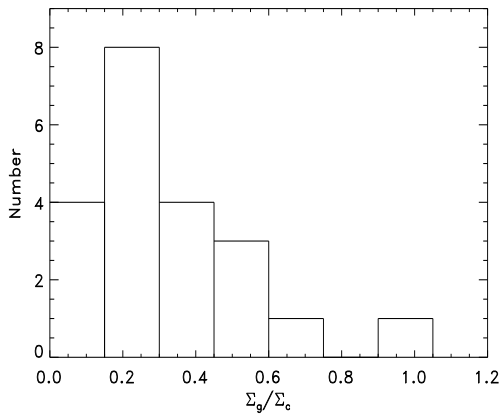}
 \caption{Histogram of the values of $\Sigma_{g}/\Sigma_{c}$ at the radius where
SFR$=10^{-4}$M$_{\odot}$yr$^{-1}$kpc$^{-2}$ in the $21$ galaxies with all
the required data presented in Leroy et al.~(2008).
}
 \label{fig:histogram}
 \end{figure}

However, none of these predictions is correct. 
To illustrate this, Figure \ref{fig:SF} shows
the radial distribution for the azimuthal averaged H$\alpha$ emission 
from Hunter et al.~(1998), and
the star formation rate, as derived by Leroy et al.~(2008)
and Bigiel et al.~(2010),
using far-ultraviolet emission and the $24\mu$m map. 
First of all, the current star formation of Ho II is not extraordinary (Hunter et al.~1998). 
Moreover,
the radius at which the furthest H$\alpha$ is detected, $R_{\rm H\alpha}$, is $\simeq 4.2$ kpc.
At $R_{\rm H\alpha}$, and for $c_{s}=6$ km s$^{-1}$, the
gaseous disk is still marginally unstable,
$\Sigma_{g}/\Sigma_{c}\simeq 1.0$.
We see that the gas Toomre parameter achieves the threshold
value $Q_{g}=1.4$ (or, equivalently, 
$\Sigma_{g}/\Sigma_{c}=0.7$; Kennicutt 1989), at $R=6$ kpc, where the local star 
formation rate is very low ($\sim 4\times 10^{-5}$M$_{\odot}$yr$^{-1}$kpc$^{-2}$). 
In the following, by comparing with
other galaxies, we will see that values for $\Sigma_{g}/\Sigma_{c}$ close to $1$
at $R_{\rm H\alpha}$ are abnormally high.

\begin{figure}
\includegraphics[width=80mm, height=105mm]{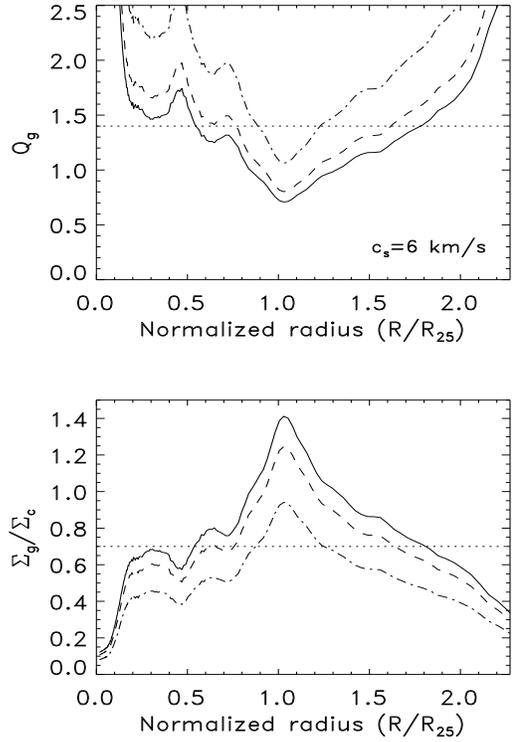}
 \caption{The $Q_{g}$-parameter (top) and $\Sigma_{g}/\Sigma_{c}$ (bottom) as a function 
of the galactocentric radius normalized by the optical radius $R_{25}$ for $D=3.4$ Mpc (solid curve),
$D=3.0$ Mpc (dashed line) and for $D=2.3$ Mpc (dot-dashed line).

}
 \label{fig:QvsD}
 \end{figure}

For a sample of nearby Sc galaxies,
Kennicutt (1989) and Martin \& Kennicutt (2001)
found that $\Sigma_{g}/\Sigma_{c}<0.75$ at $R_{\rm H\alpha}$.
Moreover, for those irregular galaxies in Hunter et al.~(1998) and
in Hunter et al.~(2011) with comfortable inclinations 
(i.e. $i>50^{\circ}$), we got that $\Sigma_{g}/\Sigma_{c}\leq 0.6$ 
at $R_{\rm H\alpha}$ (using $c_{s}=6$ km s$^{-1}$).  Finally, 
for the sample of Leroy et al.~(2008), we have calculated $\Sigma_{g}/\Sigma_{c}$
at the galactocentric radius where the star formation rate (SFR)
is $10^{-4}M_{\odot}$yr$^{-1}$kpc$^{-2}$, 
in the $21$ galaxies having all the required data (with $c_{s}=6$ km s$^{-1}$).
The histogram of the values of $\Sigma_{g}/\Sigma_{c}$ is shown in Figure
\ref{fig:histogram}. We found a mean value for $\Sigma_{g}/\Sigma_{c}$ of $0.33$.
More specifically,
$66\%$ of the galaxies have $\Sigma_{g}/\Sigma_{c}<0.35$. Only IC 2574 and HoII
have $\Sigma_{g}/\Sigma_{c}>0.6$.
Here we must note that IC 2574 presents a solid-like rotation curve. 
If the star formation rate depends on the rate of collisions between clouds,
as suggested by Tam (2000), then galaxies with solid-rotation could
exhibit larger values of $\Sigma_{g}/\Sigma_{c}$ because of their low level of shear and hence 
a lower rate of collisions between clouds. The level of shear, defined as
$|d\ln\Omega/d\ln R|$, is remarkably smaller in IC 2574; 
at the radius where SFR$=10^{-4}M_{\odot}$yr$^{-1}$kpc$^{-2}$, the level of shear
is $\sim 0.8$ in Ho II, whereas it is $\sim 0.6$ in the case of IC 2574.

In summary, the value of $\Sigma_{g}/\Sigma_{c}$ at the outer parts in Ho II,
from $3.5$ to $6$ kpc, is larger than 
the values inferred for other spiral and dwarf irregular galaxies.
This fact led us to explore the possibility that the epicyclic frequency $\kappa$
and thus the critical density $\Sigma_{c}$
have been underestimated. This may be possible if 
Ho II is either more face-on or closer (or both) than adopted.

\begin{figure}
\includegraphics[width=85mm, height=65mm]{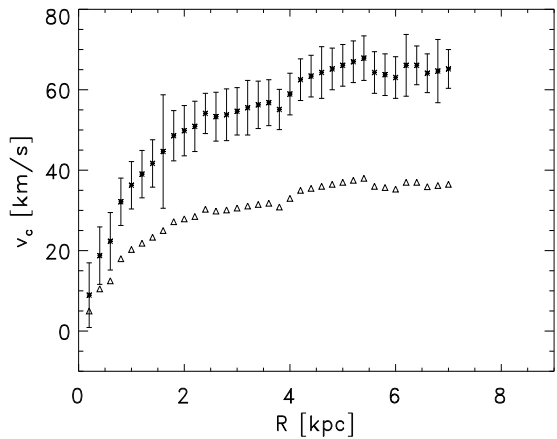}
 \caption{H\,{\sc i} rotation curve of Ho II for two different mean inclination angles: 
$\left<i\right>=49^{\circ}$
from Oh et al.~(2011) (triangles) and for $\left<i\right>=27^{\circ}$ (asterisks with error bars).
}
 \label{fig:rotation}
 \end{figure}

\begin{figure}
\includegraphics[width=80mm, height=65mm]{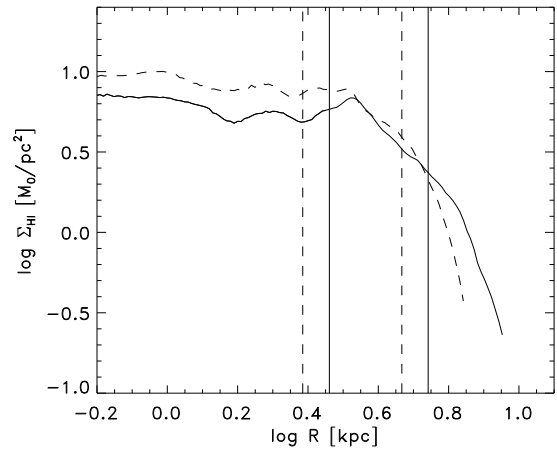}
 \caption{Radial distribution of the azimuthally averaged H\,{\sc i} surface density 
for a position angle of  $175^{\circ}$, for $\left<i\right>=49^{\circ}$ (solid line)
and for $\left<i\right>=27^{\circ}$ (dashed line). The vertical lines correspond to the radii $R_{1}$
and $R_{2}$, which depends slightly on the adopted inclination. Solid line stands for
$\left<i\right>=49^{\circ}$ and dashed lines for $\left<i\right>=27^{\circ}$; 
left  vertical lines indicate $R_{1}$ and right vertical lines $R_{2}$. 
The projected surface density was taken from THINGS (Walter et al.~2008).  
}
 \label{fig:HIdensity}
 \end{figure}

\section{Setting Ho II to `normality'}
\label{sec:normality}
In order to derive $Q_{g}$, we have assumed a distance of $D=3.4$ Mpc, 
and an inclination of $\left<i\right>=49^{\circ}$.  The distance of Ho II is probably
between $3.0$ Mpc and $3.5$ Mpc (see \S \ref{sec:properties}). 
If the galaxy distance is shorter than the adopted value, it will result in higher $Q_{g}$ 
values at a given angular radius $\alpha=R/D$
because, whereas $\Sigma_{g}(\alpha)$ is independent of $D$, the epicyclic
frequency varies $\kappa(\alpha) \propto D^{-1}$.
Figure \ref{fig:QvsD} shows the Toomre parameter and the ratio between gas surface density 
and the critical surface density for $D=3.0$ Mpc. The enhancement of the stability
parameter is not significant. In order to have $\Sigma_{g}/\Sigma_{c}=0.7$ at $R_{\rm H\alpha}$,
a value of $D=2.3$ Mpc is required. As this distance is very unlikely (the distance
uncertainty is $\sim 20\%$), we conclude that the desired level of
stability cannot be achieved by a reasonable adjustment of galaxy's
distance alone.

\begin{figure}
\includegraphics[width=85mm, height=110mm]{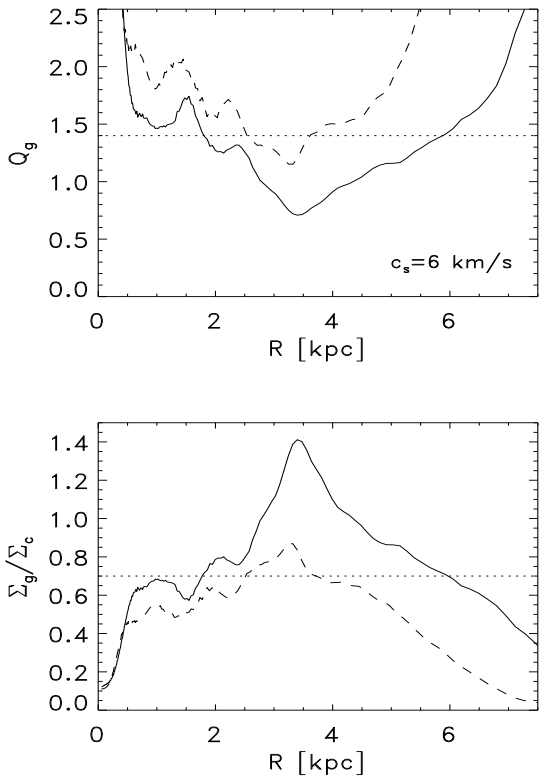}
 \caption{Radial profile of the gas Toomre parameter (top panel)
and $\Sigma_{g}/\Sigma_{c}$ (bottom panel) in Ho II for $c_{s}=6$ km s$^{-1}$,
$D=3.4$ Mpc and $\left<i\right>=27^{\circ}$ (dashed line). For comparison,
we also plot the curves for $\left<i\right>=49^{\circ}$ (solid line). For reference, 
the horizontal dotted curve corresponds to
$Q_{g}=1.4$ or, equivalently, $\Sigma_{g}/\Sigma_{c}=0.7$.

}
 \label{fig:Qinclination25}
 \end{figure}

The derived Toomre parameter profile also depends on the adopted inclination
angle of the galaxy because both $\Sigma_{g}(R)$ and $\kappa(R)$ depends
on the inclination. In particular, the amplitude of the rotation
curve and thereby $\kappa$ change as $1/\sin i$.
We have considered a model
in which the inclination angle varies slightly with galactocentric
radius from $30^{\circ}$
in the inner $2$ kpc to $24^{\circ}$ in outer parts, having
a mean inclination of Ho II of $27^{\circ}$.
In Figures \ref{fig:rotation} and \ref{fig:HIdensity}, we compare the rotation curves and the 
azimuthally-averaged H\,{\sc i} surface densities of Ho II for $D=3.4$ Mpc
and two different inclinations; 
$\left<i\right>=49^{\circ}$ and $\left<i\right>=27^{\circ}$. The resultant $Q_{g}(R)$ profiles
are shown in Figure \ref{fig:Qinclination25}, adopting a constant velocity dispersion of
$c_{s}=6$ km s$^{-1}$.  As expected, 
the derived stability parameter of the galaxy
is higher (more stable) for lower inclination angles. 

We see that if the mean inclination angle of Ho II is $27^{\circ}$ (and $D=3.4$ Mpc), then
the $\Sigma_{g}/\Sigma_{c}$-ratio at $R_{\rm H\alpha}$ takes a comfortable
value of $0.65$. This value is similar 
to those obtained in other dwarf irregular galaxies and Sc galaxies (see \S \ref{sec:newcase}).
For this inclination, $Q_{g}$ varies between a minimum of $1.2$ and
a maximum value of $1.6$ in the range\footnote{These radii
were defined in Section \ref{sec:newcase} as the radii containing
$25$ and $75$ percent of the total neutral hydrogen mass.} $R_{1}<R<R_{2}$,
and $\left<Q_{g}\right>=1.35$.
The radial variation of $Q_{g}$ between $R_{1}$ and $R_{2}$
is smaller for $\left<i\right>=27^{\circ}$ than it is when a mean inclination 
of $49^{\circ}$ is adopted
(for $\left<i\right>=49^{\circ}$, its minimum value is $0.7$ and its maximum value is $1.3$). 
Therefore, a mean inclination of $27^{\circ}$ is more consistent with the Quirk hypothesis
(1972) that self-regulation leads to a constant value of $Q_{g}$.
For reference, note that
in the sample of $21$ galaxies studied in Meurer et al.~(2013), 
$\log\left<Q_{g}\right>$ has typical rms values of only $\sim 0.05$ (see their figure 4).
Our first conclusion is that reasonable values of stability are obtained for 
a mean inclination $\sim 27^{\circ}$.
As a result, Ho II would have a rotation velocity of $\sim 60$ km s$^{-1}$
at the outer regions ($R=4$-$7$ kpc).

\begin{figure}
\includegraphics[width=80mm, height=65mm]{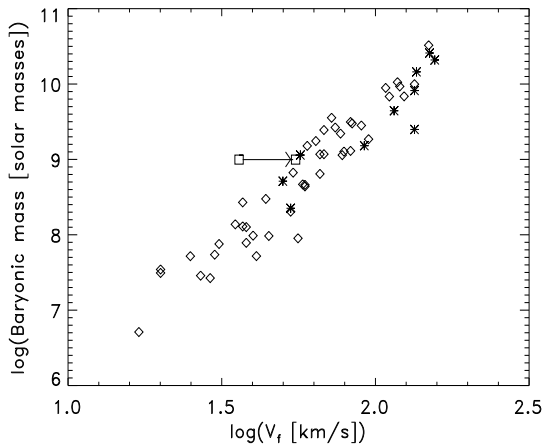}
 \caption{Baryonic Tully-Fisher relation for McGaugh (2012) sample of galaxies (diamonds)
plus the dwarf irregular galaxies in Leroy et al.~(2008) sample (asterisks). The position
of Ho II is indicated with a square for $\left<i\right>=49^{\circ}$ (left square) and for 
$\left<i\right>=27^{\circ}$ (right square).

}
 \label{fig:BTFR}
 \end{figure}

An inclination of $\sim 27^{\circ}$, required to place Ho II disk above the stability
threshold, will change the position of Ho II 
in those diagrams involving the asymptotic circular speed, such as
the baryonic Tully-Fisher relationship. 
Figure \ref{fig:BTFR} shows the baryonic Tully-Fisher relation for the sample of galaxies described
in McGaugh (2012), together with the dwarf irregular galaxies described in Leroy et al.~(2008). 
  For a mean inclination angle of $49^{\circ}$, Ho II is well outside the
baryonic Tully-Fisher relation, whereas a mean inclination of $27^{\circ}$ places Ho II
within the scatter band.

\begin{figure}
\includegraphics[width=80mm, height=65mm]{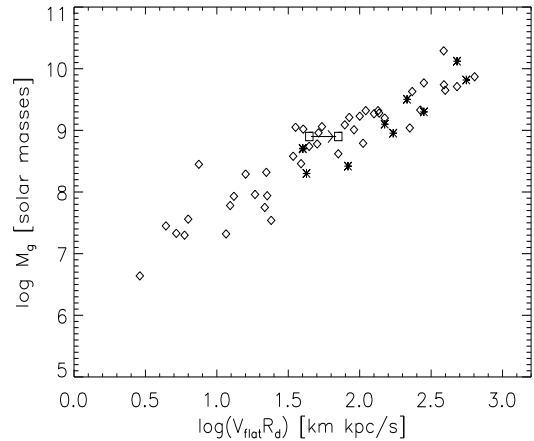}
 \caption{Total mass of gas versus $V_{\rm flat}R_{d}$ (where $V_{\rm flat}$ is the asymptotic
flat velocity and $R_{d}$ the exponential radial scale length) for McGaugh (2012)
galaxies (diamonds) and the dwarf irregular galaxies in Leroy et al.~(2008) (asterisks).
The change in the position of Ho II in this diagram, when the
inclination angle varies from $\left<i\right>=49^{\circ}$
to $\left<i\right>=27^{\circ}$, is indicated with squares.
}
 \label{fig:ZSR}
 \end{figure}

Zasov \& Smirnova (2005) found a tight correlation between the total
H\,{\sc i} mass of the galaxies and $R_{d}V_{\rm lmp}$, where $R_{d}$ is the
radial scalelength of the stellar disk and $V_{\rm lmp}$ is the circular velocity
at the last measured point\footnote{A related relationship is discussed
in Lelli et al.~(2014).}. We wish to check that the advocated mean inclination
of $27^{\circ}$ is not in conflict with this diagram. 
Figure \ref{fig:ZSR} displays the total mass gas (neutral
gas plus molecular hydrogen) versus $R_{d}V_{\rm lmp}$ for the same sample
of galaxies as used in Fig.~\ref{fig:BTFR}. We see that, although this diagram cannot
be used as a diagnostic for the inclination angle, a value 
$\left<i\right>=27^{\circ}$ keeps Ho II within the scatter in this diagram.

\begin{figure}
\includegraphics[width=80mm, height=65mm]{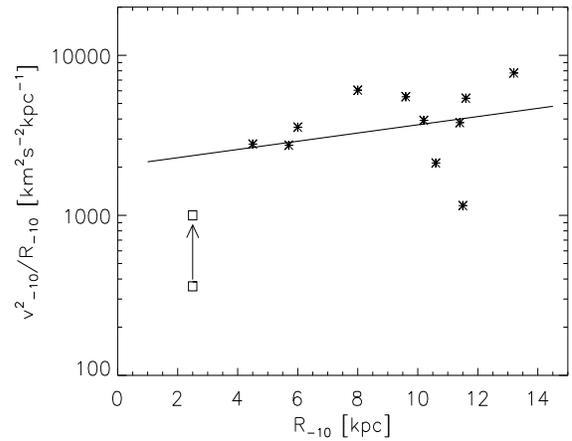}
 \caption{ Radial gravitational acceleration at $R_{-10}$ versus $R_{-10}$ for
those galaxies in Leroy et al.~(2008) sample (asterisks) having all the required data
and a flattened rotation curve. The position
of Ho II is indicated with a square for $\left<i\right>=49^{\circ}$ (lower square) and for 
$\left<i\right>=27^{\circ}$
(upper square). To guide the eye, a linear fit to the points was drawn (solid line).

}
 \label{fig:trend_SFE}
 \end{figure}

Some authors have investigated a possible correlation between the disk-average
star formation efficiency (SFE) and the orbital timescale $\tau_{\rm orb}$, defined
as $2\pi R/v_{c}$ (e.g., Kennicutt 1998).
A larger circular velocity implies a shorter orbital period.
If such a correlation exists then one could constrain the circular velocity,
and thereby the inclination of a galaxy, from observations of the SFE.
The question that arises is:
Is the SFE of Ho II more consistent with a flat circular velocity of $36$ km s$^{-1}$
or with a flat circular velocity of $60$ km s$^{-1}$?
To answer this question,
we have computed  the radial gravitational acceleration $v_{c}^{2}/R$ at the radii 
where SFE$=10^{-10}$ yr$^{-1}$, denoted by $R_{-10}$ (this occurs in the outer parts of disks),
for the galaxies in the sample of Leroy et al.~(2008),
excluding those galaxies with a solid-like rotation (i.e.~DDO 154, IC 2574, NGC 2976) and
the galaxy Ho I because its rotation curve is not flat.
Figure~\ref{fig:trend_SFE} shows the radial gravitational acceleration vs $R_{-10}$.
We see that when $\left<i\right>=49^{\circ}$,
Ho II is well outside the trend in that diagram, implying that the circular velocity at
$R_{-10}$ is too low, as compared to other galaxies.  However, an inclination angle
of $27^{\circ}$ puts Ho II back to normality.

In summary, the stability of the gaseous disk of Ho II, 
the Tully-Fisher relation, the Zasov \& Smirnova relation, and the present-day star 
formation efficiency at the outer disk, are all consistent
with an  H\,{\sc i} inclination angle of Ho II (at galactocentric radii
$3-5$ kpc) not significantly larger than $27^{\circ}$, which implies that
the amplitude of the rotation curve is $\sim 60$ km s$^{-1}$. 

All the computations above have
been derived using $D=3.4$ Mpc. If we assume a distance of $3.0$ Mpc,
an inclination angle $\sim 32^{\circ}$ is needed to have the required level
of stability of the gaseous disk. Thus, uncertainties in the distance
have a minor role.

\section{Ho II: implications for MOND}
\label{sec:MONDcase}
Milgrom (1983) proposed a modification of Newton  gravitation law to explain
the dynamics of galaxies without invoking dark matter. S\'anchez-Salcedo et al.~(2013)
provided a sample of gas-rich dwarf galaxies that could be problematic for MOND.
One of those galaxies was Ho II. S\'anchez-Salcedo et al.~(2013) argued that the rotation
curve of Ho II is consistent with MOND only if the H\,{\sc i} inclination is $\sim 25^{\circ}$.
In the previous sections, we have provided indirect evidence suggesting that,  in fact,
the inclination of Ho II is possibly close to $25^{\circ}$ (see also Gentile et al.~2012).
With this new inclination, the amplitude of the rotation curve of Ho II can be matched under MOND.

One of the arguments in favor of an inclination angle of $\sim 25^{\circ}$ 
was based on stability analysis.
For significantly higher inclination angles, Ho II would be abnormally unstable, as compared to 
other galaxies.
The level of stability of galaxy, i.e.~the Toomre parameter, was studied under Newtonian gravity.
However, it depends on the adopted gravitational law. In the following, we will compute the
Toomre parameter of the gaseous disk under the MOND framework, $Q_{g,M}$, 
to see if its level of stability is also reasonable.

The MOND version of Poisson's equation is given by:
\begin{equation}
{\mbox{\boldmath $\nabla$}}\cdot 
\left[\mu\left(\frac{|{\mbox{\boldmath $\nabla$}}\Phi|}{a_{0}}\right)
{\mbox{\boldmath $\nabla$}}\Phi\right] =4\pi G \rho,
\end{equation}
where $\rho$ is the density distribution, $\Phi$ the gravitational
potential, $a_{0}$ is a universal acceleration of the order of 
$10^{-8}$ cm s$^{-2}$, and $\mu(x)$ is some interpolating function with
the property that $\mu(x)=x$ for $x\ll 1$ and $\mu(x)=1$ for $x\gg 1$ (Bekenstein \& Milgrom 1984).
Milgrom (1989) demonstrated that the Toomre parameter in MOND
is:
\begin{equation}
Q_{M}=\mu^{+}(1+L^{+})^{1/2} Q_{N},
\label{eq:QM}
\end{equation}
where $Q_{N}$ is the Newtonian Toomre parameter as defined in 
Eq.~(\ref{eq:Qnewton}),  
$L\equiv d\ln\mu/d\ln x$ is the logarithmic derivate of $\mu$, whereas
$\mu^{+}$ and $L^{+}$ are the values of $\mu$ and $L$
just above the disk.

The two most popular choices for the interpolating function are
the ``simple'' $\mu$-function, suggested by Famaey \& Binney (2005),
\begin{equation}
\mu(x)=\frac{x}{1+x},
\end{equation}
and the ``standard'' $\mu$-function
\begin{equation}
\mu(x)=\frac{x}{\sqrt{1+x^{2}}},
\end{equation}
proposed by Milgrom (1983).
Since we are interested in Ho II dwarf galaxy whose dynamics
lies in the deep MOND regime (that is, $x=g/a_{0}\ll 1$), our results 
are not sensitive to the exact form of the interpolating function.
We will use the simple $\mu$-function with 
$a_{0}=1.2\times 10^{-8}$ cm s$^{-2}$.

In the case of the simple $\mu$-function, the Toomre parameter
in MOND differs from the Toomre parameter
in Newton by the factor $\mu^{+}(1+L^{+})^{1/2}=
\mu^{+}(2-\mu^{+})^{1/2}$. Since this factor is $\leq 1$, 
the Toomre parameter in MOND is lower than it is in
the corresponding Newtonian galaxy with a dark halo.

\begin{figure}
  \includegraphics[width=80mm,height=75mm]{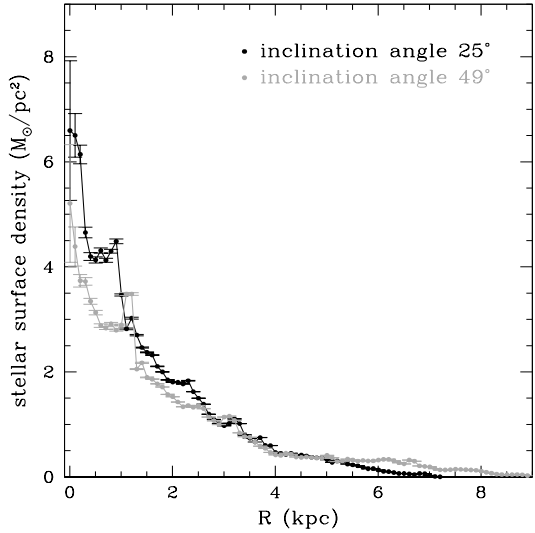}
\vskip 0.4cm
  \caption{Azimuthally averaged stellar surface density against radius from the
galaxy center for two angle inclinations $25^{\circ}$ and $49^{\circ}$, using
the ZCR method.}
  \label{fig:stellarML}
\end{figure}

We have computed $Q_{g,M}$ for our model with $\left<i\right>=27^{\circ}$.
This inclination is required in order to match the amplitude of the 
rotation curve with the value predicted by MOND. 
To derive $\mu^{+}$ and $L^{+}$, we assumed that the disk
is very thin so that the vertical Newtonian gravitational field just
above and just below the disk has a magnitude 
$2\pi G (\Sigma_{g}+\Sigma_{\star})$.
In order to compute the stellar surface density $\Sigma_{\star}$, 
we derived the pixel-to-pixel stellar mass-to-light ratio applying the method
proposed by Zibetti, Charlot \& Rix (2009; hereafter ZCR) to the B, V and R images
from SINGS (Kennicutt et al.~2003). The azimuthally averaged stellar surface density,
using the ZCR stellar mass-to-light ratio ($\Upsilon_{\star,\rm ZCR}$),
is shown in Figure \ref{fig:stellarML}. For completeness and in order to show its
dependence on the adopted inclination, we plot the stellar surface density for two 
different inclinations ($25^{\circ}$ and $49^{\circ}$).

\begin{figure}
  \includegraphics[width=80mm,height=105mm]{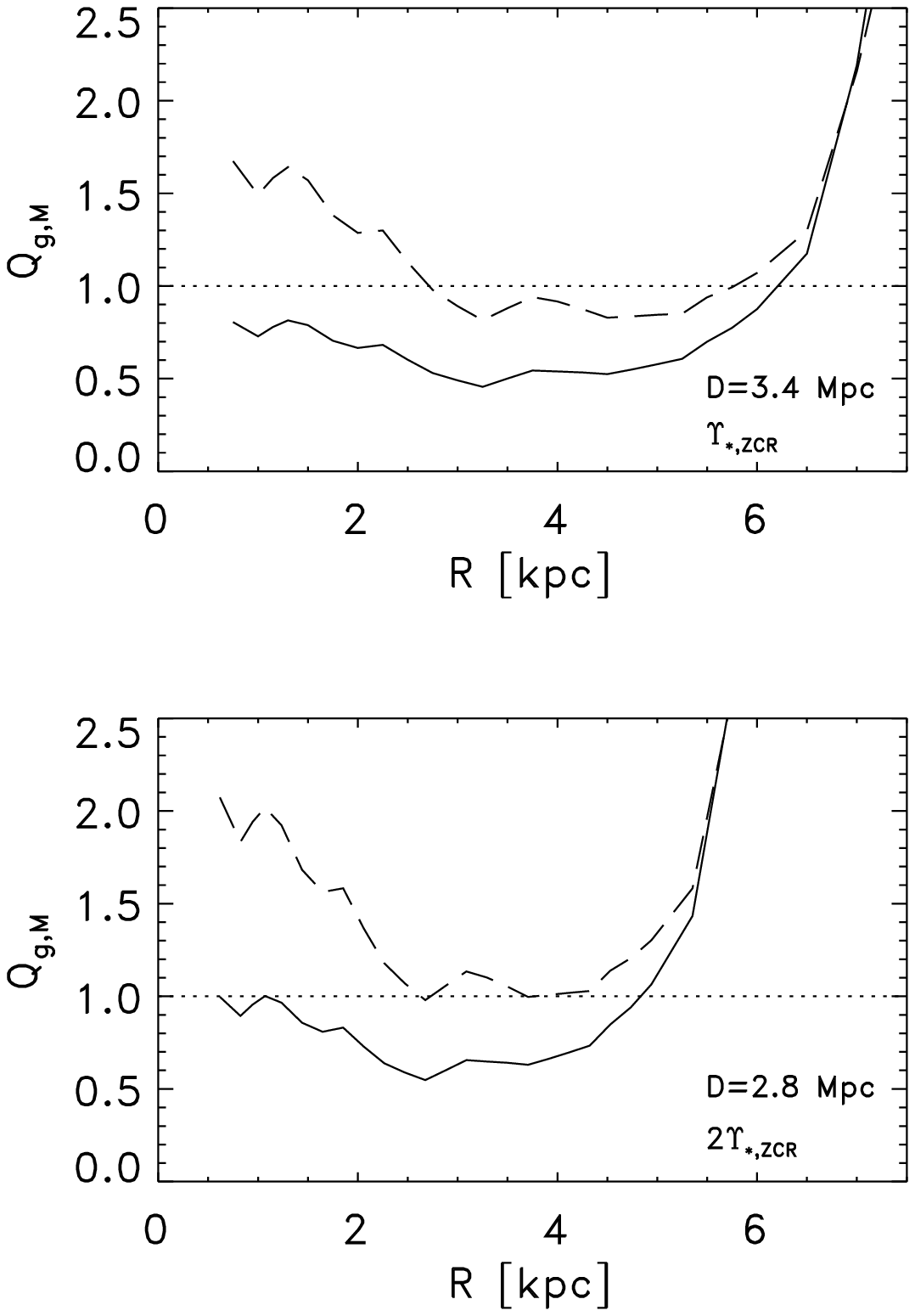}
\vskip 0.4cm
  \caption{
Radial distribution of the Toomre parameter in
MOND using a constant gas velocity dispersion
of $6$ km s$^{-1}$ (solid line) and using the velocity dispersion
from the second-moment map (dashed line). We have used the simple
$\mu$-function.  In the top panel, we have assumed $D=3.4$ Mpc and
the $\Upsilon_{\star}$-value derived using the ZCR method. In the 
bottom panel,  we used $D=2.8$ Mpc and twice the $\Upsilon_{\star,\rm ZCR}$-value. 
The gas surface density $\Sigma_{g}$ was taken as $1.4$ times 
the azimuthally averaged H\,{\sc i}
surface density. In order to estimate the gas surface density, a mean inclination 
angle of $27^{\circ}$ was used.
  }
  \label{fig:MONDToomre}
\end{figure}

Figure \ref{fig:MONDToomre} shows $Q_{g,M}$ in Ho II,
as a function of radius, using two different choices for $c_{s}$:
(1) a constant value $c_{s}=6$ km s$^{-1}$
(solid line) and (2) the observed H\,{\sc i} velocity dispersion from the
second-moment map (dashed line). 
Consider first the Toomre parameter derived 
using $\Upsilon_{\star,\rm ZCR}$ and $D=3.4$ Mpc (top panel).
What we see in Figure \ref{fig:MONDToomre} is that, when
$c_{s}=6$ km s$^{-1}$, the MOND Toomre parameter is less than $1$
 at any radius between $R=1$ and $R=6$ kpc. We find no correlation
between the Toomre parameter, derived assuming that $c_{s}$ is
constant, and the star formation
rate (see Fig.~\ref{fig:SF}). 
If the critical surface density is a real threshold for star formation
in MOND, then we would expect, following the same reasoning as in
\S \ref{sec:newcase}, a huge star formation activity throughout the disk
of Ho II up to a radius of $6$ kpc. However, this is not observed.
If, instead of a constant sound speed, we take the observed H\,{\sc i}
velocity dispersion from the second-moment map, we find no correlation
between star formation activity and the Toomre parameter either.
More importantly, we find that, adopting the observed velocity dispersion, the gas 
Toomre parameter $Q_{g,M}$ is higher than $1$, as required
for stability, within $2.5$ kpc, but it is nearby $1$ in a significant portion of the disk
(3 kpc $<R<6$ kpc), implying that it is marginally unstable and thereby
very responsive to perturbations. 

The values of $Q_{g,M}$ are not very sensitive to reasonable changes in the
adopted distance and stellar mass-to-light ratio (see Figure \ref{fig:MONDToomre}). 
The effect of adopting a different $D$ can be computed as follows. 
If the stellar mass-to-light
ratio is given, the Newtonian gravitational acceleration $v_{c,N}^{2}/R$ at a certain
angular radius does not
depend on the adopted distance. Thus,  $\mu$ does not depend on $D$. Therefore,
$Q_{g,M}$ depends only on $D$ through $\kappa$, which scales as $\propto D^{-1/2}$.
This implies that a reduction in the adopted distance from $3.4$ Mpc to $2.8$ Mpc 
increases the values of $Q_{g,M}$ by $10\%$. On the other hand, increasing 
the stellar mass-to-light ratio leads to a higher $Q_{g,M}$
because the circular velocity and $\kappa$ will both increase. This effect is however small
because the stellar mass is only $20\%$ of the mass in gas, even when a value of $2\Upsilon_{\star,\rm ZCR}$ is
assumed.
Figure \ref{fig:MONDToomre} shows $Q_{g,M}$ when the distance is reduced by $20\%$ and
the stellar mass-to-light ratio is augmented by a factor of $2$. The gas stability parameter in this case
is slightly higher than it is for the referecence values but it is still close to unity at the interval
$2.5$-$4.5$ kpc in radius.

S\'anchez-Salcedo \& Hidalgo-G\'amez (1999) already noticed that
the dwarf irregular galaxies IC 2574 and NGC 1560
have $Q_{g,M}<1$ along a significant portion of the disk.
What stabilizes MONDian gas-rich dwarf galaxies 
against the formation of grand-design spiral arms and/or widespread
star formation up to the radius where $Q_{g,M}\sim 1.4$?
We urge MOND simulators to consider the problem of the stability of gas-rich
dwarf galaxies as it may provide a key test for MOND.

\section{Final remarks and conclusions}
\label{sec:conclusions}
The existence of an intrinsic scatter in the baryonic Tully-Fisher relation may have
strong implications in galaxy evolution theories and modified gravity.
To establish any scatter, one needs 
precise determinations of the amplitude of rotation curves 
and thus good estimates of the
inclination of the galaxies, especially for galaxies with low inclinations. 
It is very difficult to associate a confidence
interval to the inclination from tilted-ring analysis. 
Oh et al.~(2011) presented a comprehensive analysis of the
mass distributions of seven dwarf
galaxies, being especially careful on the effect of
inclination on the rotation curves. Regarding Ho II,
they claimed that ``{\it all ring parameters are well determined...}''
and argued that ``{\it as can be seen from not only the tilted-ring
analysis but also the comparison of rotation velocities in the Appendix
it is unlikely that Ho II has an inclination ($\sim 25^{\circ}$) as low
as that inferred from the baryonic Tully-Fisher relation}''. If so, Ho II would not satisfy
the baryonic Tully-Fisher relation and would be problematic for MOND (S\'anchez-Salcedo
et al.~2013). Motivated by this result,
Gentile et al.~(2012) re-analyzed the H\,{\sc i} data cube and 
found that the inclination
may be much closer to face-on than previously derived (see also McGaugh 2011).
It is clear that the main source of systematic uncertainties in
the determination of the amplitude of the rotation curve in Ho II
is its inclination.

An amplitude of the rotation curve lower than
expected could be caused by an overestimate of its inclination angle in the sky.
In this work,
we provide some indirect but robust evidence, based on stability
arguments, to suggest that the inclination
of Ho II is smaller than the value inferred by tilted-ring analysis.
We have derived the mean inclination of Ho II in order 
to place it above the stability threshold at the radius of $\sim 4$ kpc, where the
SFE is as low as $2\times 10^{-11}$ yr$^{-1}$
(Leroy et al.~2008).  We have shown that, under Newtonian
gravity, this occurs for a mean inclination of $\sim 27^{\circ}$. 
This inclination puts Ho II back into the baryonic
Tully-Fisher relation.

One could ask what is the impact of this new inclination
in the position of Ho II in other diagrams.
We find that if $\left<i\right>=27^{\circ}$, Ho II lies in the 
right location in the $M_{g}$ vs $V_{\rm flat}R_{d}$ trend. 
We also find that the radial gravitational acceleration $v_{c}^{2}/R$ at the
radius where SFE$= 10^{-10}$ yr$^{-1}$ is too low in Ho II as compared
to the values in other galaxies for inclination angles of $\sim 50^{\circ}$, 
but it is within the scatter when $\left<i\right>=27^{\circ}$. 

All the above arguments indicate that Ho II will be back to normality 
provided that the mean inclination angle is $\sim 27^{\circ}$. This inclination
is also consistent with the ellipticity of the isocontours of the total H\,{\sc i}
surface density map, in the outer parts of the galaxy (Gentile et al.~2012).

It is important to note that adopting this inclination, 
the rotation curve of Ho II is in agreement with MOND (Gentile et al.~2012;
S\'anchez-Salcedo et al.~2013). However, a similar stability analysis under
MOND indicates that the gas Toomre
parameter, calculated using the observed velocity dispersion,
is close to $1$ between $0.8R_{25}$ and $1.8R_{25}$ in galactocentric radius.
To be satisfactory, MOND should explain why gravitational instabilities 
do not promote a burst of star formation or a grand-design
spiral arm with widespread star formation at the outer parts of Ho II, i.e.~between
$0.8R_{25}$ and $1.8R_{25}$. 
Indeed, we were unable to find 
a possible connection between the degree of
large-scale gravitational instability and
the locations of star formation. 
MONDian numerical simulations are needed to understand why gas-rich dwarf galaxies, 
such as Ho II or IC 2574, appear unperturbed even when the level of self-gravity of 
their disks is so important.

\acknowledgements
The authors are indebted to the referee for a thoughtful report.
The authors made use of THINGS
`The H\,{\sc i} Nearby Galaxy Survey' (Walter et al.~2008).
This work was supported by the following projects: CONACyT 
165584, PAPIIT IN106212 and SIP-20141379.

{}

\end{document}